\documentstyle[12pt,epsf,rotating]{article}
\def\fo{\hbox{{1}\kern-.25em\hbox{l}}}

\newcommand{\newc}{\newcommand}

\newc{\lcal}{\int {\cal L}dt}

\newc{\LSP}{{\chi^0_1}}
\newc{\stauR}{{\tilde \tau_R}}
\newc{\stau}{{\tilde \tau_1}}
\newc{\mstop}{m_{\tilde{t}}}
\newc{\mHpm}{m_{H^\pm}}
\newc{\gsim}{\lower.7ex\hbox{$\;\stackrel{\textstyle>}{\sim}\;$}}
\newc{\lsim}{\lower.7ex\hbox{$\;\stackrel{\textstyle<}{\sim}\;$}}
\newc{\ie}{{\it i.e.}}
\newc{\etal}{{\it et al.}}
\newc{\eg}{{\it e.g.}}
\newc{\kev}{\hbox{\rm\,keV}}
\newc{\mev}{\hbox{\rm\,MeV}}
\newc{\gev}{\hbox{\rm\,GeV}}
\newc{\tev}{\hbox{\rm\,TeV}}
\newc{\xpb}{\hbox{\rm\, pb}}
\newc{\xfb}{\hbox{\rm\, fb}}

\newc{\mtop}{m_t}
\newc{\mbot}{m_b}
\newc{\mz}{m_Z}
\newc{\mw}{M_W}
\newc{\alphasmz}{\alpha_s(m_Z^2)}
\newc{\swsq}{\sin^2\theta_W}
\newc{\tw}{\tan\theta_W}
\newc{\cw}{\cos\theta_W}
\newc{\sw}{\sin\theta_W}
\newc{\BR}{\hbox{\rm BR}}
\newc{\zbb}{Z\to b\bar}
\newc{\Gb}{\Gamma (Z\to b\bar b)}
\newc{\Gh}{\Gamma (Z\to \hbox{\rm hadrons})}
\newc{\rbsm}{R_b^\hbox{\rm sm}}
\newc{\rbsusy}{R_b^\hbox{\rm susy}}
\newc{\drb}{\delta R_b}

\newc{\sgn}{\mbox{sgn}}

\newc{\tbeta}{\tan\beta}
\newc{\uL}{{\tilde u_L}}
\newc{\uR}{{\tilde u_R}}
\newc{\cL}{{\tilde c_L}}
\newc{\cR}{{\tilde c_R}}
\newc{\tL}{{\tilde t_L}}
\newc{\tR}{{\tilde t_R}}
\newc{\dL}{{\tilde d_L}}
\newc{\dR}{{\tilde d_R}}
\newc{\sL}{{\tilde s_L}}
\newc{\sR}{{\tilde s_R}}
\newc{\bL}{{\tilde b_L}}
\newc{\bR}{{\tilde b_R}}
\newc{\eL}{{\tilde e_L}}
\newc{\eR}{{\tilde e_R}}
\newc{\mhp}{m_{H^\pm}}
\newc{\mhalf}{m_{1/2}}
\newc{\emt}{{e/\mu /\tau}}

\newc{\lR}{\tilde{l}_R}
\newc{\lL}{\tilde{l}_L}
\newc{\nL}{\tilde{\nu}_L}
\newc{\na}{\chi^0_1}
\newc{\nb}{\chi^0_2}
\newc{\nc}{\chi^0_3}
\newc{\nd}{\chi^0_4}
\newc{\ca}{\chi^{\pm}_1}
\newc{\cb}{\chi^{\pm}_2}
\newc{\camp}{\chi^\mp_1}
\newc{\cbmp}{\chi^\mp_1}
\newc{\capos}{\chi^{+}_1}
\newc{\caneg}{\chi^{-}_1}
\newc{\phit}{\phi_t}
\newc{\phib}{\phi_b}
\newc{\phiew}{\phi_{ew}}
\newc{\htz}{h^0_t}
\newc{\hbz}{h^0_b}
\newc{\hewz}{h^0_{ew}}
\newc{\hsmz}{h^0_{sm}}
\newc{\huz}{h^0_u}
\newc{\hsusyz}{h^0_{susy}}

\def\mp{M_P}

\def\beq{\begin{equation}}
\def\eeq{\end{equation}}
\def\bea{\begin{eqnarray}}
\def\eea{\end{eqnarray}}
\def\slashchar#1{\setbox0=\hbox{$#1$}           
   \dimen0=\wd0                                 
   \setbox1=\hbox{/} \dimen1=\wd1               
   \ifdim\dimen0>\dimen1                        
      \rlap{\hbox to \dimen0{\hfil/\hfil}}      
      #1                                        
   \else                                        
      \rlap{\hbox to \dimen1{\hfil$#1$\hfil}}   
      /                                         
   \fi}                                         %
\catcode`@=11
\long\def\@caption#1[#2]#3{\par\addcontentsline{\csname
  ext@#1\endcsname}{#1}{\protect\numberline{\csname
  the#1\endcsname}{\ignorespaces #2}}\begingroup
    \small
    \@parboxrestore
    \@makecaption{\csname fnum@#1\endcsname}{\ignorespaces #3}\par
  \endgroup}
\catcode`@=12

\begin{document}

\baselineskip=18pt

\begin{titlepage}

\begin{center}
\vspace{1cm}

{\Large \bf Effects of Supersymmetric Threshold Corrections on
High-Scale Flavor Textures}

\vspace{0.5cm}

{\bf{Altan\, \c{C}ak{\i}r$^a$,  Levent\, Solmaz$^{a,b}$}}

\vspace{.8cm}

{\it $^a$Department of Physics, Izmir Institute of Technology, IZTECH, Turkey,\\ TR35430}

{\it $^{b}$Department of Physics, Bal{\i}kesir University, Bal{\i}kesir, Turkey,\\ TR10100}

\end{center}
 \vspace{1cm}

\begin{abstract}
\medskip
Integration of  superpartners out of the spectrum induces
potentially large contributions to Yukawa couplings. These
corrections, the supersymmetric threshold corrections, therefore
influence the CKM matrix prediction in a non-trivial way. We study
effects of threshold corrections on high-scale flavor structures
specified at the gauge coupling unification scale in
supersymmetry. In our analysis, we first consider high-scale
Yukawa textures which qualify phenomenologically viable at tree
level, and find that they get completely disqualified after
incorporating the threshold corrections. Next, we consider Yukawa
couplings, such as those with five texture zeroes, which are
incapable of explaining flavor-changing proceses. Incorporation of
threshold corrections, however, makes them phenomenologically
viable textures. Therefore, supersymmetric threshold corrections
are found to leave observable impact on Yukawa couplings of
quarks, and any confrontation of high-scale textures with
experiments at the weak scale must take into account such
corrections.

\end{abstract}
\bigskip
\end{titlepage}

\tableofcontents
\section{Introduction and Motivation}
Supersymmetric theories with general soft breaking terms possess a
number of flavor and CP violation sources \cite{soft}. In general,
they are nested in the rigid and soft sectors of the theory, and
bear no correlation or selection rules whatsoever. This is the
case in supergravity and superstring scenarios where K{\"a}hler
metric and superpotential couplings are all generic matrices in
the space of fermion flavors and depend on the compactification
scheme employed. This rather high degree of freedom in sources,
textures and structures of  the flavor mixings at GUT/string scale
needs to be refined by confronting them with experimental data
especially on rare processes. In general, testing high-scale
flavor structures with experimental data involves three basic
ingredients:
\begin{enumerate}
\item Specification of flavor textures in rigid and soft sectors
at the messenger scale (which we take to be the MSSM gauge
coupling unification scale $Q= M_{GUT} \sim 10^{16}\ {\rm GeV}$).

\item Rescaling of lagrangian parameters to low-scale $Q=M_{weak}\sim {\rm TeV}$ via renormalization group flow.
This stage is particularly important due to ($i$) largeness of the logs ($\log M_{GUT}/M_{weak}$) involved, and
($ii$) modifications of flavor structures because of mixings with others.

\item Integration out of the superpartners at $M_{weak}$ to
achieve an effective theory which comprises the SM particle
spectrum with possible imprints of supersymmety in various
couplings. For FCNC phenomenology this step is important as it
induces flavor-nonuniversal couplings of gauge and Higgs bosons to
fermions.
\end{enumerate}
Any high-scale flavor structure specified in step 1 is classified
to be phenomenologically viable if it agrees with experimental
data after step 3. The first two steps have been widely discussed
in literature by identifying flavor violation sources in general
supergravity \cite{sugra,sugra1} and confronting them with
experimental data on fermion masses and mixings as well as various
observables in kaon and beauty systems \cite{sugra1,fcnc}.

So far analysis of the third step above has been restricted to
${\rm TeV}$-scale supersymmetry where gauge \cite{gauge} and Higgs
\cite{higgs} bosons have been found to develop flavor-changing
couplings to fermions. In particular, emphasis has been put on the
couplings of $Z$ \cite{gauge} and Higgs \cite{higgsafter} to $b
\overline{s}$ since mixing between second and third generation
fermions exhibits a theoretically clean and experimentally wide
room for new physics. These analyses have led to conclusion that
flavor violation sources in  sfermion sector can have a big impact
on Higgs phenomenology as well as various rare processes in kaon
and beauty systems \cite{higgs}.

It is thus of prime phenomenological interest to know what impact
the integration-out of sparticles can leave on high-scale flavor
textures below $M_{weak}$. Stating more specifically, can
integrating sparticles out of the spectrum render a given
high-scale otherwise-viable texture inappropriate or generate CKM
matrix from solely the soft sector or modify effects of Yukawa
couplings on the soft sector? These are some of the questions
which will be addressed in the present work.

In Sec. 2 below we briefly discuss the formalism for determining
effects of supersymmetric threshold corrections. We mainly follow
results of \cite{higgs} therein. In Sec. 3,  we first discuss in
Sec.3.1 sensitivities of the GUT-scale CKM-ruled, hierarchic and
democratic Yukawa textures to supersymmetric threshold corrections
when trilinear couplings are proportional to Yukawas. In Sec.3.2
we investigate effects of flavor mixings in squark mass-squared
matrices on textures analyzed in Sec. 3.1. In Sec.3.3, we
determine effects of threshold corrections on Yukawa textures
which would not qualify physical at tree level. In Sec. 4 we
conclude.

\section{The Formalism}
The superpotential of the MSSM
\begin{eqnarray}
\label{rigid} \widehat{W} = \widehat{U} {\bf Y_u} \widehat{Q}
\widehat{H}_u + \widehat{D} {\bf Y_d} \widehat{Q} \widehat{H}_d +
\widehat{E} {\bf Y_e} \widehat{L} \widehat{H}_d + \mu
\widehat{H}_u \widehat{H}_d
\end{eqnarray}
encodes the rigid parameters $\mu$ and Yukawa couplings ${\bf
Y_{u,d,e}}$ (of up quarks, down quarks and of leptons) each being
a $3\times 3$ non-hermitian matrix in the space of fermion
flavors.

The breakdown of supersymmetry is parameterized by a set of soft
($i.e.$ operators of dimension $\leq 3$) terms \cite{soft}
\begin{eqnarray}
\label{soft}
 {\cal{L}}_{soft} &=&m_{H_u}^2 H_u^{\dagger} H_u + m_{H_d}^2 H_d^{\dagger} H_d +
\widetilde{Q}^{\dagger} {\bf m_Q^2} \widetilde{Q} + \widetilde{U}
{\bf m_U^2} \widetilde{U}^{\dagger} + \widetilde{D} {\bf m_D^2}
\widetilde{D}^{\dagger} + \widetilde{L}^{\dagger} {\bf m_L^2}
\widetilde{L} + \widetilde{E} {\bf m_E^2}
\widetilde{E}^{\dagger}\nonumber\\
&+& \left[\widetilde{U} {\bf Y_u^A} \widetilde{Q} {H}_u +
\widetilde{D} {\bf Y_d^A} \widetilde{Q} {H}_d + \widetilde{E} {\bf
Y_e^A} \widetilde{L} {H}_d + \mu B {H}_u {H}_d +
\frac{1}{2}\sum_{\alpha} M_{\alpha} \lambda_{\alpha}
\lambda_{\alpha} + \mbox{h.c.}\right]
\end{eqnarray}
where trilinear couplings ${\bf Y_{u,d,e}^A}$ like Yukawas
themselves are non-hermitian flavor matrices whereas the sfermion
mass-squareds ${\bf m_{Q,\dots,E}^2}$ are all hermitian. In
general, all of the parameters in the second line and off-diagonal
entries of the sfermion mass-squared matrices are endowed with
CP--odd phases; they serve as sources of CP violation beyond the
SM. The Yukawa matrices, trilinear couplings and sfermion
mass-squareds facilitate flavor violation in processes mediated by
sparticle loops. The MSSM possesses 21 mass parameters, 36 mixing
angles and 40 CP-odd phases in addition to ones in the SM
\cite{dimo}. Consequently, there is a 97-dimensional parameter
space to be scanned in confronting theory with experiments at
$M_{weak}$. In supergravity or string models the parameters of
(\ref{rigid}) and (\ref{soft}) are determined by compactification
mechanism and structure of the internal manifold \cite{sugra}.

The parameters of (\ref{rigid}) and (\ref{soft}) are scale-
dependent. They are rescaled to $Q=M_{weak}$ via the MSSM RGEs
\cite{rge} with boundary conditions specified at $Q=M_{GUT}$. The
RG running of model parameters is crucial. In fact, various
phenomena central to supersymmetry phenomenology $e.g.$ gauge
coupling unification, radiative electroweak breaking, induction of
flavor structures even for flavor-blind soft terms are pure
renormalization effects. The Yukawa couplings, $\mu$ parameter and
gauge couplings form a coupled closed set of observables
\cite{closed} in that their scale dependencies are not affected by
soft-breaking sector unless some sparticles are decoupled before
reaching $M_{weak}$. On the other hand, running of the soft masses
depend explicitly on rigid parameters of the theory, and they
develop, among other things, novel flavor structures thanks to the
Yukawa matrices. For instance, evolution of the soft mass-squared
of left-handed squarks
\begin{eqnarray}
\label{rg-mq}  \frac{d {\bf m_Q^2}}{d t} &=& {\bf m_Q^2}
\left({\bf Y_u}^{\dagger} {\bf Y_u} + {\bf Y_d}^{\dagger} {\bf
Y_d}\right) + \left({\bf Y_u}^{\dagger} {\bf Y_u} + {\bf
Y_d}^{\dagger} {\bf Y_d}\right) {\bf m_Q^2}\nonumber\\  &+& 2
\left( {\bf Y_u}^{\dagger} {\bf m_U^2} {\bf Y_u} + {\bf
Y_d}^{\dagger} {\bf m_D^2} {\bf Y_d} + {\bf Y_u^A}^{\dagger} {\bf
Y_u^A} +  {\bf Y_d^A}^{\dagger} {\bf Y_d^A}\right)\nonumber\\
&+& 2 \left(m_{H_u}^2 {\bf Y_u}^{\dagger} {\bf Y_u}  + m_{H_d}^2
{\bf Y_d}^{\dagger} {\bf Y_d}\right)\nonumber\\ &-& 2
\left(\frac{16}{3} g_3^2 \left|M_3\right|^2 + 3 g_2^2 \left|
M_2\right|^2 + \frac{1}{15} g_1^2 \left|M_1\right|^2\right) {\bf
1}\,,
\end{eqnarray}
with $t \equiv (4 \pi)^{-2} \log Q/M_{GUT}$, shows explicitly how
flavor structure of a given parameter, say ${\bf m_Q^2}$, at a
given scale $Q$ senses those of the remaining parameters. Indeed,
flavor mixings exhibited by ${\bf m_Q^2}$ at $Q= M_{weak}$ can
stem from ${\bf m_{Q,U,D}^2}$ or ${\bf Y_{u,d}}$ or ${\bf
Y_{u,d}^A}$ or all of them. Therefore, a given pattern of flavor
mixings in, for instance, kaon system can be sourced by various
flavor matrices in rigid as well as soft sectors of the theory.

The flavor structures at $M_{weak}$ arising from solutions of RGEs
are further rehabilitated by taking into account the decoupling of
sparticles at the supersymmetric threshold. Indeed, once part of
the sparticles are integrated out of the spectrum the effective
theory below $M_{weak}$ can exhibit sizeable non-standard effects
in certain scattering channels of the SM particles
\cite{higgs,higgsafter,gauge}. Taking the effective theory below
$M_{weak}$ to be two-Higgs-doublet model (2HDM) one finds
\begin{eqnarray}
\label{effYukawa} {\bf Y_d}^{eff} &=& {\bf Y_d}(M_{weak}) -
\gamma^{d} + \tan \beta\,
\Gamma^{d}\nonumber\\
{\bf Y_u}^{eff} &=& {\bf Y_u}(M_{weak}) + \gamma^{u} - \cot
\beta\, \Gamma^{u}
\end{eqnarray}
where ${\bf Y_{d,u}}(M_{weak})$ are solutions of the corresponding
RGEs evaluated  at $Q=M_{weak}$, and  $\gamma^{d,u}$ and
$\Gamma^{d,u}$ are flavor matrices arising from squark-gluino and
squark-Higgsino loops. Their explicit expressions can be found in
\cite{higgs}.

The physical quark fields
are obtained by rotating the original gauge eigenstate fields via
the unitary matrices $V_{R,L}^{u,d}$ that diagonalize ${\bf
Y_{u,d}}^{eff}$:
\begin{eqnarray}
\label{diag-yukawa} \left(V_{R}^{d}\right)^{\dagger} {\bf
Y_d}^{eff} V_{L}^{d} = \overline{\bf Y_{d}} \;,\;\;
\left(V_{R}^{u}\right)^{\dagger} {\bf Y_u}^{eff} V_{L}^{u} =
\overline{\bf Y_{u}}
\end{eqnarray}
where $\overline{\bf Y_{d}}= \mbox{diag.}\left(\overline{h_d},
\overline{h_s}, \overline{h_b}\right)$ and $\overline{\bf Y_{u}}=
\mbox{diag.}\left(\overline{h_u}, \overline{h_c},
\overline{h_t}\right)$ are physical Yukawa  matrices whose entries
are directly related to running quark masses at $Q=M_{weak}$:
\begin{eqnarray}
\overline{h_u} = \frac{g_{2}(M_{weak})\,m_{u}(M_{weak})}{\sqrt{2} M_W \sin\beta}\;,\;\; \overline{h_d} =
\frac{g_{2}(M_{weak})\,m_{d}(M_{weak})}{\sqrt{2} M_W \cos\beta}
\end{eqnarray}
with similar expressions for other generations.


In general, whatever flavor textures are adopted at $M_{GUT}$, the
resulting CKM matrix, $V_{CKM}^{corr} \equiv
\left(V_L^u\right)^{\dagger}\,V_L^d$, must agree with the existing
experimental bounds \cite{pdg}. Clearly, in the limit of vanishing
threshold corrections $\Gamma^{u,d}$ and $\gamma^{u,d}$, physical
CKM matrix $V_{CKM}^{corr}$ reduces to $V_{CKM}^{tree}$ computed
by diagonalizing ${\bf Y_{u,d}}(M_{weak})$ . Reiterating, it is
with comparison of the predicted CKM matrix, $V_{CKM}^{corr}$,
with experiment that one can tell if a high-scale texture,
classified to be viable at tree-level by considering
$V_{CKM}^{tree}$ only, is spoiled by the supersymmetric threshold
corrections. The experimental bounds on the absolute magnitudes of
the CKM entries (at $90 \%$ CL) read collectively as:
\begin{eqnarray}
\label{exp-ckm} \left|V_{CKM}^{exp}\right|=
\left(\begin{array}{ccc}
\begin{tabular}{|c|c|}
  \hline 0.9739 & 0.9751 \\ \hline
\end{tabular}
& \begin{tabular}{|c|c|}
  \hline 0.2210 & 0.2270\\ \hline
\end{tabular}
& \begin{tabular}{|c|c|}
  \hline 0.0029& 0.0045 \\ \hline
\end{tabular}\\ \\
\begin{tabular}{|c|c|}
 \hline  0.2210 & 0.2270 \\ \hline
\end{tabular} & \begin{tabular}{|c|c|}
 \hline  0.9730 & 0.9744 \\ \hline
\end{tabular} & \begin{tabular}{|c|c|}
 \hline  0.0390 & 0.0440 \\ \hline
\end{tabular} \\ \\
\begin{tabular}{|c|c|}
 \hline  0.0048 & 0.0140 \\ \hline
\end{tabular} & \begin{tabular}{|c|c|}
 \hline  0.0370 & 0.0430 \\ \hline
\end{tabular} & \begin{tabular}{|c|c|}
  \hline 0.9990 & 0.9992\\ \hline
\end{tabular}
\end{array} \right)
\end{eqnarray}
where left (right) window of $\begin{tabular}{|c|c|}
 \hline  { }& { }\\ \hline
\end{tabular}$ in each entry refers to lower (upper) experimental
bound on the associated CKM element. Clearly, the largest
uncertainity occurs in $|V_{td}|$. These matrix elements are
measured at $Q=M_{Z}$, and for a comparison with predictions of
the effective theory below $Q=M_{weak}$ they have to be scaled
from $M_Z$ up to $M_{weak}$. This can be done without having a
detailed knowledge of the particle spectrum of the effective 2HDM
at $M_{weak}$ ( as emphasized above, the effective theory may
consist of some light superpartners in which case beta functions
of certain couplings get modified as exemplified by analyses of
$b\rightarrow s \gamma$ decay in effective supersymmetry
\cite{bsgam}) since RG running of the CKM elements is such that
$V_{CKM}(1,1)$, $V_{CKM}(1,2)$, $V_{CKM}(2,1)$, $V_{CKM}(2,2)$ and
$V_{CKM}(3,3)$ do not evolve with energy scale, to an excellent
approximation \cite{pokorski}. Therefore, it is rather safe to
confront the CKM matrix predicted by the effective theory at
$M_{weak}$ with the experimental results (\ref{exp-ckm}) entry by
entry excluding, however, $V_{CKM}(1,3)$, $V_{CKM}(3,1)$,
$V_{CKM}(2,3)$ and $V_{CKM}(3,2)$ for which renormalization
effects can be sizeable.

In the next section, we will compute supersymmetric threshold
corrections to Yukawa couplings of quarks for certain prototype
flavor textures defined at $Q=M_{GUT}$. In particular, we will
evaluate radiatively corrected CKM matrix as well as couplings of
the Higgs bosons to quarks to determine the impact of the
decoupling of squarks out of the spectrum at $M_{weak}$ on
scattering processes at energies accessible to present and future
colliders.

\section{High-Scale Textures and Threshold Corrections}

First of all, for standardization and easy comparison with
literature ($e.g.$ with the computer codes ISAJET \cite{isajet}
and SOFTSUSY \cite{allanach}) we take SPS1a$^{\prime}$ conventions
for supersymmetric parameters \cite{sps1a}
\begin{eqnarray}
\tan\beta=10\;,\; m_{0}=70\ \rm{GeV}\;,\; A_0=-300\ \rm{GeV}\;,\;
m_{1/2}=250\ \rm{GeV}
\end{eqnarray}
and completely neglect supersymmetric CP-violating phases, as
mentioned before.

Instead of scanning a 97-dimensional parameter space for
specifying what high-scale parameter ranges are useful for what
low-energy observables, which is actually what has to be done, we
simplify the analysis by focussing on certain prototype textures
at high scale. In general, for any flavor matrix in any sector of
the theory there exist, boldly speaking, three extremes: ($i$)
completely diagonal, ($ii$) hierarchical, and ($iii$) democratic
textures. There are, of course, a continuous infinity of textures
among these extremes; however, for definiteness and clarity in our
analysis we will focus on these three structures.

\subsection{Flavor violation from Yukawas and trilinear couplings}
In this subsection we investigate effects of superymmetric
threshold corrections on high-scale textures in which Yukawa
couplings exhibit non-trivial flavor mixings and so do the
trilinear couplings since we take
\begin{eqnarray}
\label{YApropY} {\bf Y_{u,d,e}^A} = A_0 {\bf Y_{u,d,e}}
\end{eqnarray}
at the GUT scale. The soft mass-squareds, on the other hand, are
taken entirely flavor conserving $i.e.$ they are strictly diagonal
and universal at the GUT scale. It is with direct proportionality
of trilinear couplings with Yukawas and certain ansatze for Yukawa
textures that, we will study below sensitivities of certain
high-scale Yukawa structures to supersymmetric threshold
corrections at the ${\rm TeV}$ scale.

\subsubsection{CKM-ruled texture}
We take Yukawa couplings of up and down quarks to be
\begin{eqnarray}
\label{minfv-yukawa}
{\bf Y_{u}}&=&\rm{diag }\left(3.5\ 10^{-6},1.3\ 10^{-3},0.4566\right)\nonumber\\
{\bf Y_{d}}&=& \left(
\begin{array}{ccc}
 {6.2368\ 10^{-5}} & -{1.4272\ 10^{-5}} &
 {5.9315\ 10^{-7}\ e^{0.3146 i}} \\
 {2.4640\ 10^{-4}} &
   {1.07074\ 10^{-3}} &
   -{4.0458\ 10^{-5}} \\
   1.6495\ 10^{-4}\ e^{1.047 i}&
   {1.81465\ 10^{-3}} & {4.8476\ 10^{-2}}
\end{array} \right)
\end{eqnarray}
with no flavor violation in the lepton sector: ${\bf Y_e}= \mbox{diag.}\left(1.9\ 10^{-5}, 4\ 10^{-3},
0.071\right)$. The flavor violation effects are entirely encoded in ${\bf Y_d}$ which exhibits a CKM-ruled
hierarchy in similarity to Yukawa textures analyzed in \cite{sugra1} $i.e.$ this choice of boundary values of
the Yukawas leads to correct CKM matrix \cite{pdg} at $M_{weak}$ upon integration of the RGEs.

At the weak scale the Yukawa matrices, trilinear couplings and
squark soft mass-squareds serve as sources of flavor violation.
The trilinear couplings, under two-loop RG running \cite{rge} with
boundary conditions (\ref{YApropY}), attain the flavor structures
\begin{eqnarray}
\label{minfv-YA} {\bf Y}_{u}^A&=&\left( \begin{array}{c c c}\matrix{ \ -7.2 \ 10^{-3}& 0 & 0 \cr 1.70\
{10}^{-6}\ e^{0.5641 i} & -2.67 & 2.9 \ 10^{-4} \cr \ 6.24\ e^{1.047 i} \  10^{-3} & 6.8 \ 10^{-2} & -532.7 \cr
} \
\end{array}\right)\nonumber\\
{\bf Y}_{d}^A&=&\left( \begin{array}{c c c}\matrix{ \ -0.204 & -0.191 &-0.138\ e^{-1.039 i} \cr -0.567 & -3.495
& -1.436 \cr -0.384\ e^{1.046 i} & -4.19 & -134.24 \cr  } \
\end{array}\right)~\end{eqnarray}
both measured in ${\rm GeV}$ at $M_{weak}=1\ {\rm TeV}$. Clearly,
${\bf Y_u^{A}}$ is essentially diagonal whereas $(2,3)$, ($3,2$)
and $(2,2)$ entries of ${\bf Y_d^{A}}$ are of the same size.

Though they start with completely diagonal and universal boundary
values, the squark soft squared masses develop flavor-changing
entries at $M_{weak}=1\ {\rm TeV}$:
\begin{eqnarray}
\label{minfv-msq} {\bf m_Q^2}&=& \left(533.67\ {\rm GeV}\right)^{2} \left(\begin{array}{ccc}
  1.07 & 0.0 & 0.0 \\
   0.0& 1.07 & -2.2\ 10^{-4} \\
  0.0& -2.2\ 10^{-4} &0.86
\end{array}\right)
\nonumber\\
{\bf m_D^2}&=& \left(530.76\ {\rm GeV}\right)^{2} \left(\begin{array}{ccc}
  1.01 & 0.0 & 0.0 \\
   0.0& 1.01 & -1.5\ 10^{-4} \\
  0.0& -1.5\ 10^{-5} & 0.99
\end{array}\right)
 \end{eqnarray}
with ${\bf m_U^2}= \left(497.11\ {\rm GeV}\right)^{2}\ \mbox{diag.}\left(1.15, 1.15, 0.69\right)$. The numerical
values of the parameters above exhibit good agreement with well-known codes like ISAJET \cite{isajet} and
SOFTSUSY \cite{allanach}.

The presence of flavor violation in the soft sector of the
low-energy theory gives rise to non-trivial corrections to Yukawa
couplings and in turn to the CKM matrix. Indeed, use of
(\ref{minfv-YA}) and (\ref{minfv-msq}) in \cite{higgs} introduces
certain corrections to the tree-level Yukawa matrices ${\bf
Y_{u,d}}(M_{weak})$ to generate  ${\bf Y_{u,d}^{eff}}$ in
(\ref{effYukawa}). In fact, $V_{CKM}^{tree}$ (obtained from ${\bf
Y_{u,d}}(M_{weak})$) and $V_{CKM}^{corr}$ (obtained from ${\bf
Y_{u,d}}^{eff}$) compare to exhibit spectacular differences:
\begin{eqnarray}
\label{minfv-CKM}  \begin{tabular}{|c|c|}
  \hline $\left|V_{CKM}^{tree}\right|$ & $\left|V_{CKM}^{corr}\right|$ \\ \hline
\end{tabular} =
\left(
\begin{array}{ccc}
  \begin{tabular}{|c|c|}
   \hline $0.9746$ & $0.9795$ \\ \hline
  \end{tabular} & \begin{tabular}{|c|c|}
    \hline 0.2241 & 0.2015 \\ \hline
  \end{tabular} & \begin{tabular}{|c|c|}
    \hline 0.0037 & 0.0034 \\ \hline
  \end{tabular} \\ \\
  \begin{tabular}{|c|c|}
    \hline 0.2240 & 0.2014 \\ \hline
  \end{tabular} & \begin{tabular}{|c|c|}
    \hline $0.9737$ & $0.9788$ \\ \hline
  \end{tabular} & \begin{tabular}{|c|c|}
   \hline 0.0406 & 0.0375 \\ \hline
  \end{tabular} \\ \\
  \begin{tabular}{|c|c|}
   \hline $0.0079$ & $0.0066$ \\ \hline
  \end{tabular} & \begin{tabular}{|c|c|}
    \hline $0.0400$ &  $0.0371$\\ \hline
  \end{tabular} & \begin{tabular}{|c|c|}
    \hline $0.99917$ & $0.9993$ \\ \hline
  \end{tabular}
\end{array}
\right)
\end{eqnarray}
where left (right) window of $\begin{tabular}{|c|c|}
 \hline  { }& { }\\ \hline
\end{tabular}$
in $(i,j)$-th entry refers to $\left|V_{CKM}^{tree}(i,j)\right|$ (
$\left|V_{CKM}^{corr}(i,j)\right|$). Clearly, $|V_{CKM}^{tree}|$
agrees very well with $\left|V_{CKM}^{exp}\right|$ in
(\ref{exp-ckm}) entry by entry. This qualifies
(\ref{minfv-yukawa}) to be the correct high-scale texture given
experimental FCNC bounds at $Q=M_{Z}$. However, radiative
corrections induced by decoupling of squarks, gluinos and
Higgsinos at the supersymmetric threshold $M_{weak}=1\ {\rm TeV}$
is seen to leave a rather strong impact on the CKM entries.
Consider for instance $(1,1)$ entries of $V_{CKM}^{exp}$,
$V_{CKM}^{tree}$ and $V_{CKM}^{corr}$. Present experiments provide
a $1.64 \sigma$ significance to $\left|V_{CKM}^{exp}(1,1)\right|$
around a mean value of $0.745$ as is seen from (\ref{exp-ckm}).
The tree-level prediction, $\left|V_{CKM}^{tree}(1,1)\right|$,
takes the value of $0.9746$ which is rather close to the center of
the experimental interval. However, once supersymmetric threshold
corrections are included this tree-level prediction gets modified
to $\left|V_{CKM}^{corr}(1,1)\right|= 0.9795$. This value is
obviously far beyond the existing experimental limits as it is a
$13.39 \sigma$ effect. Similarly, $
\left|V_{CKM}^{corr}(1,2)\right|$, $
\left|V_{CKM}^{corr}(2,1)\right|$, $
\left|V_{CKM}^{corr}(2,2)\right|$ and $
\left|V_{CKM}^{corr}(3,3)\right|$ are, respectively, $12.36
\sigma$, $12.36 \sigma$, $11.95 \sigma$ and $2.30 \sigma$ effects.
 Obviously, deviation of $\left|V_{CKM}^{corr}(i,j)\right|$ from
$\left|V_{CKM}^{tree}(i,j)\right|$ (comparison with experiments at
$Q=M_Z$ is meaningful especially for  $(i,j)=(1,1), (1,2), (2,1),
(3,3)$ entries whose scale dependencies are known to be rather
mild \cite{pokorski}), when the latter falls well inside the
experimentally allowed range, obviously violates existing
experimental bounds in (\ref{exp-ckm}) by several standard
deviations. Consequently, supersymmetric threshold corrections
entirely disqualify the high-scale texture (\ref{minfv-yukawa})
being the correct texture to reproduce the FCNC measurements at
the weak scale. This case study, based on numerical values for
Yukawa entries in (\ref{minfv-yukawa}), manifestly shows the
impact of supersymmetric threshold corrections on high-scale
textures which qualify viable at tree level.

The physical quark fields, which arise after the unitary rotations
(\ref{diag-yukawa}), acquire the masses
\begin{eqnarray}
\overline{\bf M_{u}}(M_{weak})= \mbox{diag.}\left(\simeq 0, 0.545,
149.45\right)\,,\; \overline{\bf M_{d}}(M_{weak})=
\mbox{diag.}\left(3.35\ 10^{-3}, 5.76\ 10^{-2}, 2.33\right)
\end{eqnarray}
all measured in ${\rm GeV}$. In this physical basis for quark
fields, $V_{CKM}^{corr}$ governs the strength of charged current
vertices for each pair of up and down quarks. These mass
predictions are to be evolved down to $Q=M_{Z}$ to make
comparisons with experimental results. This evolution depends on
the effective theory below $M_{weak}$. Speaking conversely, the
high-scale texture (\ref{minfv-yukawa}) has to be folded in such a
way that resulting mass and mixing patterns for quarks agree with
experiments below the sparticle threshold $M_{weak}$.

\subsubsection{Hierarchical texture}
The Yukawa couplings are taken to have the structure (as can be
motivated from \cite{lavignac})
\begin{eqnarray}
\label{hierfv-yukawa} {\bf Y_{u}}&=& \left(\begin{array}{ccc}
  2.6463\ 10^{-4} & 5.8163\ 10^{-4} i & - 1.0049\ 10^{-2} \\
  - 5.8163\ 10^{-4} i & 2.2587\ 10^{-3} & 1.0049\ 10^{-5} i \\
  -4.8233\ 10^{-3} & -9.0437\ 10^{-6} i & 0.495
\end{array} \right)\nonumber\\
{\bf Y_{d}}&=& \left(
\begin{array}{ccc}
 {3.9808\ 10^{-4}} & {8.1167\ 10^{-4}\ e^{0.734 i}} &
 {-1.1431\ 10^{-3}} \\
  {8.1167\ 10^{-4}\ \ e^{- 0.734 i}} &
   {2.7997\ 10^{-3}} &
   {2.04844\ 10^{-3}} i \\
-1.1431\ 10^{-3} &
   - {1.6461\ 10^{-3}} i & {4.97\ 10^{-2}}
\end{array} \right)
\end{eqnarray}
with no flavor violation in the lepton sector: ${\bf Y_e}=
\mbox{diag.}\left(1.9\ 10^{-5}, 0.004, 0.071\right)$. Here both
${\bf Y_u}$ and ${\bf Y_d}$ exhibit a hierarchically organized
pattern of entries .  In a sense, the hierarchic nature of ${\bf
Y_d}$ in (\ref{minfv-yukawa}) is now extended to ${\bf Y_{u}}$ so
as to form a complete hierarchic pattern for quark Yukawas at the
GUT scale.

At the weak scale, the Yukawa matrices above, trilinear couplings,
and squark soft mass-squareds serve as sources of flavor
violation. The trilinear couplings, under two-loop RG running
\cite{rge} with boundary conditions (\ref{YApropY}), obtain the
flavor structures
\begin{eqnarray}
\label{hierfv-YA} {\bf Y}_{u}^A&=& \left( \begin{array}{c c c}\matrix{ \ -0.4315 & - 1.1442 i & 10.637 \cr
1.1466 i & -4.4531 & -4.8631\ 10^{-3} i \cr \ 5.0657 & -0.1046 i & -524.07 \cr } \
\end{array}\right)\nonumber\\
{\bf Y}_{d}^A&=&\left( \begin{array}{c c c}\matrix{ \ -1.2934& -2.6494\ e^{0.734 i} & 3.1221 \cr  2.6428\
e^{-0.731 i} & -9.1395 & 5.2606 i \cr  3.4532 & -5.6827 i & -135.861 \cr  } \
\end{array}\right)~\end{eqnarray}
both measured in ${\rm GeV}$ at $M_{weak}=1\ {\rm TeV}$. Clearly,
in contrast to (\ref{minfv-YA}), now both ${\bf Y_u^{A}}$ and
${\bf Y_d^A}$ develop sizeable off-diagonal entries, as expected
from (\ref{hierfv-yukawa}).

Though they start with completely diagonal and universal boundary
values, the squark soft squared masses develop flavor-changing
entries at $M_{weak}=1\ {\rm TeV}$:
\begin{eqnarray}
\label{hierfv-msq} {\bf m_Q^2}&=& \left(533.69\ {\rm GeV}\right)^{2} \left(\begin{array}{ccc}
  1.07 & 1.9\ 10^{-5}\ e^{1.144 i} & 2.14\ 10^{-3} \\
   1.9\ 10^{-5}\ e^{-1.144 i}  & 1.07 & 3.17\ 10^{-4} i \\
  2.14\ 10^{-3} & -3.17\ 10^{-4} i &0.86
\end{array}\right)
\nonumber\\
{\bf m_U^2}&=& \left(496.76\ {\rm GeV}\right)^{2} \left(\begin{array}{ccc}
  1.16 & - 6.66\ 10^{-6} i & 9.6\ 10^{-3} \\
  6.66\ 10^{-6} i   & 1.16 & -1.4\ 10^{-5} i \\
  9.6\ 10^{-3} &  1.4\ 10^{-5} i &0.685
\end{array}\right)
\nonumber\\
{\bf m_D^2}&=& \left(531.07\ {\rm GeV}\right)^{2} \left(\begin{array}{ccc}
  1.01 & 3.3\ 10^{-5} \ e^{1.06 i}& 3.75\ 10^{-4} \\
   3.3\ 10^{-5} \ e^{- 1.06 i}& 1.01 & -6.62\ 10^{-4} i\\
  3.75\ 10^{-4}& 6.62\ 10^{-4} i & 0.99
\end{array}\right)
 \end{eqnarray}
whose average values show good agreement with (\ref{minfv-msq})
but certain off-diagonal entries exhibit significant enhancements
when the corresponding entries of Yukawas and trilinear couplings
are sizeable.

The flavor-violating entries of Yukawas, trilinear couplings and
soft mass-squareds collectively generate radiative contributions
$\gamma^{u,d}$, $\Gamma^{u,d}$ to the Yukawa couplings below
$M_{weak}$ \cite{higgs}. In fact, $V_{CKM}^{tree}$ (obtained from
${\bf Y_{u,d}}(M_{weak})$) and $V_{CKM}^{corr}$ (obtained from
${\bf Y_{u,d}}^{eff}$) confront as follows:
\begin{eqnarray}
\label{hierfv-CKM}  \begin{tabular}{|c|c|}
  \hline $\left|V_{CKM}^{tree}\right|$ & $\left|V_{CKM}^{corr}\right|$ \\ \hline
\end{tabular} =
\left(
\begin{array}{ccc}
  \begin{tabular}{|c|c|}
   \hline $0.9745$ & $0.9773$ \\ \hline
  \end{tabular} & \begin{tabular}{|c|c|}
    \hline 0.2243 & 0.2118 \\ \hline
  \end{tabular} & \begin{tabular}{|c|c|}
    \hline 0.0049 & 0.0034 \\ \hline
  \end{tabular} \\ \\
  \begin{tabular}{|c|c|}
    \hline 0.2240 & 0.2116 \\ \hline
  \end{tabular} & \begin{tabular}{|c|c|}
    \hline $0.9737$ & $0.9766$ \\ \hline
  \end{tabular} & \begin{tabular}{|c|c|}
   \hline 0.0417 & 0.0379 \\ \hline
  \end{tabular} \\ \\
  \begin{tabular}{|c|c|}
   \hline $0.0109$ & $0.0091$ \\ \hline
  \end{tabular} & \begin{tabular}{|c|c|}
    \hline $0.0405$ &  $0.0370$\\ \hline
  \end{tabular} & \begin{tabular}{|c|c|}
    \hline $0.99912$ & $0.99927$ \\ \hline
  \end{tabular}
\end{array}
\right)
\end{eqnarray}
where left (right) window of $\begin{tabular}{|c|c|}
 \hline  { }& { }\\ \hline
\end{tabular}$
in $(i,j)$-th entry refers to $\left|V_{CKM}^{tree}(i,j)\right|$ (
$\left|V_{CKM}^{corr}(i,j)\right|$). Clearly, $|V_{CKM}^{tree}|$
falls well inside the $1.64 \sigma$ experimental interval in
(\ref{exp-ckm}) entry by entry. In this sense, Yukawa matrices in
(\ref{demfv-yukawa})  qualify  to be the correct high-scale
textures given present experimental determination of $V_{CKM}$ at
$Q=M_{Z}$. However, this agreement between experiment and theory
gets spoiled strongly by the inclusion of supersymmetric threshold
corrections. Indeed, as is shown comparatively by
(\ref{demfv-CKM}), $V_{CKM}^{corr}$ violates the bounds in
(\ref{exp-ckm}) significantly. More precisely,
$\left|V_{CKM}^{corr}(1,1)\right|$, $
\left|V_{CKM}^{corr}(1,2)\right|$, $
\left|V_{CKM}^{corr}(2,1)\right|$, $
\left|V_{CKM}^{corr}(2,2)\right|$, $
\left|V_{CKM}^{corr}(3,3)\right|$ turn out to have $7.65 \sigma$,
$6.83 \sigma$, $6.77 \sigma$, $6.79 \sigma$, $3.28 \sigma$
significance levels, respectively. These significance levels are
far beyond the existing experimental $1.64 \sigma$ intervals
depicted in (\ref{exp-ckm}). As a result, supersymmetric threshold
corrections are found to entirely disqualify the high-scale
texture (\ref{hierfv-yukawa}) to be the correct texture to
reproduce the FCNC measurements at the weak scale. This case study
therefore shows the impact of supersymmetric threshold corrections
on high-scale textures which qualify viable at tree level.

The physical quark fields, which arise after the unitary rotations
(\ref{diag-yukawa}), acquire the masses
\begin{eqnarray}
\overline{\bf M_{u}}(M_{weak})= \mbox{diag.}\left(0.0065, 0.98, 153.82\right)\,,\; \overline{\bf
M_{d}}(M_{weak})= \mbox{diag.}\left(0.0071, 0.155, 2.37\right)
\end{eqnarray}
all measured in ${\rm GeV}$. In this physical basis for quark
fields, $V_{CKM}^{corr}$ is responsible for charged current
interactions in the effective theory below $M_{weak}$. The morale
of the analysis above is that, the high-scale flavor structures
(\ref{hierfv-yukawa}) are to be modified in such a way that
$V_{CKM}^{corr}$ agrees with $V_{CKM}^{exp}$ with sufficient
precision. Aftermath, the question is to predict quark masses
appropriately at $Q=M_{weak}$ so that, depending on the particle
spectrum of the effective theory beneath, existing experimental
values of quark masses at $Q=M_Z$ are reproduced correctly.

\subsubsection{Democratic texture}
In this subsection, we take Yukawa couplings to be (as can be
motivated from relevant works \cite{democratic})
\begin{eqnarray}
\label{demfv-yukawa} {\bf Y_{u}}&=& \left(\begin{array}{ccc}
  0.1475 & 0.1443 & 0.1458 \\
  0.1443 & 0.1475 & 0.1458 \\
  0.1456 & 0.1458 & 0.1456
\end{array} \right)\nonumber\\
{\bf Y_{d}}&=& \left(
\begin{array}{ccc}
 0.01583 & 0.01452 (1- 10^{-2} i) &
  0.01553 (1- 10^{-2} i) \\
   0.01452 (1+ 10^{-2} i) &
   0.01944 &  0.01617 (1+ 2\ 10^{-2} i)\\
 0.01551 (1+ 10^{-2} i) &0.01617 (1- 2\ 10^{-2} i)
    & 0.01604
\end{array} \right)
\end{eqnarray}
with no flavor violation in the lepton sector: ${\bf Y_e}= \mbox{diag.}\left(1.9\ 10^{-5}, 4\ 10^{-3},
0.071\right)$. Here both ${\bf Y_u}$ and ${\bf Y_d}$ exhibit an approximate democratic structure so that ${\bf
Y_{u,d}}(M_{weak})$ generate correctly masses and mixings of the quarks at the weak scale. Clearly, in the exact
democratic limit two of the quarks from each sector remain massless, and therefore, a realistic flavor structure
is likely to come from small perturbations of the exact democratic texture \cite{democratic}. Another important
feature of exact democratic texture is that all higher powers of Yukawas reduce to Yukawas themselves up to a
multiplicative factor, and this gives rise to linearization of and in turn direct solution of Yukawa RGEs in the
form of an RG rescaling of the GUT scale texture \cite{closed}. These properties remain approximately valid for
perturbed democratic textures like (\ref{demfv-yukawa}).

At the weak scale, the Yukawa matrices above, trilinear couplings,
and squark soft mass-squareds serve as sources of flavor
violation. The trilinear couplings, under two-loop RG running
\cite{rge} with boundary conditions (\ref{YApropY}), obtain the
flavor structures
\begin{eqnarray}
\label{demfv-YA} {\bf Y}_{u}^A&=&-\left( \begin{array}{c c c}\matrix{ \ 182.44 & 175.57 & 178.81 \cr \ 175.69 &
182.32 & 178.81\cr \ 178.62 & 178.81 & 178.67 \cr } \
\end{array}\right)\nonumber\\
{\bf Y}_{d}^A&=&-\left( \begin{array}{c c c}\matrix{ \ 44.41 & 40.07\ e^{-0.0117 i} & 43.39\ e^{0.0115 i} \cr
39.44 e^{ 0.0101i} & 55.46\ e^{-0.0013 i} & 44.82\ e^{0.0218 i} \cr 43.09\ e^{- 0.099 i} & 45.17\ e^{-0.0216 i}
& 44.79\ e^{0.0016 i} \cr }\
\end{array}\right)~\end{eqnarray}
both measured in ${\rm GeV}$ at $M_{weak}=1\ {\rm TeV}$. Though
not shown explicitly, each entry of ${\bf Y_u^A}$ is complex with
a phase around $10^{-7}$ -- $10^{-6}$ in size.

Though they start with completely diagonal and universal boundary
values, the squark soft squared masses develop flavor-changing
entries at $M_{weak}=1\ {\rm TeV}$:
\begin{eqnarray}
\label{demfv-msq} {\bf m_Q^2}&=& \left(533.67\ {\rm GeV}\right)^{2} \left(\begin{array}{ccc}
  1.0 & 0.0672 & 0.0670 \\
  0.0672 & 1.0 & 0.0673 \\
 0.0670 & 0.0673 &1.0
\end{array}\right)
\nonumber\\
{\bf m_U^2}&=& \left(497.38\ {\rm GeV}\right)^{2} \left(\begin{array}{ccc}
  1.0 & 0.1526 & 0.1524 \\
  0.1526& 1.0 &  0.1524\\
  0.1524 & 0.1524  &1.0
\end{array}\right)
\\
{\bf m_D^2}&=& \left(530.59\ {\rm GeV}\right)^{2} \left(\begin{array}{ccc}
  1.0 & 5.046\ 10^{-3} \ e^{-0.01i}&  4.826\ 10^{-3}\ e^{0.01i}\\
  5.046\ 10^{-3} \ e^{0.01i}& 1.0 & 5.289\ 10^{-3} \ e^{0.02i}\\
  4.826\ 10^{-3} \ e^{-0.01i}& 5.289\ 10^{-3} \ e^{-0.02i}& 1.0
\end{array}\right)\nonumber
 \end{eqnarray}
whose average values show good agreement with (\ref{minfv-msq})
and (\ref{hierfv-msq}). The off-diagonal entries of each squark
soft mass-squared are of similar size due to the democratic
structure of the Yukawa couplings. The flavor-mixing entries
$m_{\widetilde U}^2$ are the largest among all three mass
squareds.

The flavor-violating entries of Yukawas, trilinear couplings and
soft mass-squareds collectively generate radiative contributions
$\gamma^{u,d}$, $\Gamma^{u,d}$ to the Yukawa couplings below
$M_{weak}$ \cite{higgs}. In fact, $V_{CKM}^{tree}$ (obtained from
${\bf Y_{u,d}}(M_{weak})$) and $V_{CKM}^{corr}$ (obtained from
${\bf Y_{u,d}}^{eff}$) confront as follows:
\begin{eqnarray}
\label{demfv-CKM}  \begin{tabular}{|c|c|}
  \hline $\left|V_{CKM}^{tree}\right|$ & $\left|V_{CKM}^{corr}\right|$ \\ \hline
\end{tabular} =
\left(
\begin{array}{ccc}
  \begin{tabular}{|c|c|}
   \hline $0.9748$ & $0.9685$ \\ \hline
  \end{tabular} & \begin{tabular}{|c|c|}
    \hline 0.2229 & 0.2490 \\ \hline
  \end{tabular} & \begin{tabular}{|c|c|}
    \hline 0.0083 & 0.0085 \\ \hline
  \end{tabular} \\ \\
  \begin{tabular}{|c|c|}
    \hline 0.2229 & 0.2489 \\ \hline
  \end{tabular} & \begin{tabular}{|c|c|}
    \hline $0.9739$ & $0.9674$ \\ \hline
  \end{tabular} & \begin{tabular}{|c|c|}
   \hline 0.0421 & 0.0463 \\ \hline
  \end{tabular} \\ \\
  \begin{tabular}{|c|c|}
   \hline $0.0092$ & $0.0104$ \\ \hline
  \end{tabular} & \begin{tabular}{|c|c|}
    \hline $0.0419$ &  $0.0459$\\ \hline
  \end{tabular} & \begin{tabular}{|c|c|}
    \hline $0.99908$ & $0.99889$ \\ \hline
  \end{tabular}
\end{array}
\right)
\end{eqnarray}
where left (right) window of $\begin{tabular}{|c|c|}
 \hline  { }& { }\\ \hline
\end{tabular}$
in $(i,j)$-th entry refers to $\left|V_{CKM}^{tree}(i,j)\right|$ (
$\left|V_{CKM}^{corr}(i,j)\right|$). Obviously, $|V_{CKM}^{tree}|$
agrees very well with $\left|V_{CKM}^{exp}\right|$ in
(\ref{exp-ckm}) entry by entry. This qualifies
(\ref{demfv-yukawa}) to be the correct high-scale texture given
present experimental determination of $V_{CKM}$ at $Q=M_{Z}$. The
most striking aspect of (\ref{demfv-CKM}) is the fact that
supersymmetric threshold corrections push $V_{CKM}^{tree}$ beyond
the experimental bounds. More precisely,
$\left|V_{CKM}^{corr}(1,1)\right|$, $
\left|V_{CKM}^{corr}(1,2)\right|$, $
\left|V_{CKM}^{corr}(2,1)\right|$, $
\left|V_{CKM}^{corr}(2,2)\right|$, $
\left|V_{CKM}^{corr}(3,3)\right|$ turn out to have $17.22 \sigma$,
$14.21 \sigma$, $14.21 \sigma$, $ 15.22 \sigma$, $16.40  \sigma$
significance levels, respectively. These are obviously far beyond
the existing experimental $1.64 \sigma$ significance intervals
depicted in (\ref{exp-ckm}). As a result, supersymmetric threshold
corrections are found to entirely disqualify the high-scale
texture (\ref{demfv-yukawa}) to be the correct texture to
reproduce the FCNC measurements at the weak scale.

Here, it is worthy of noting that deviation of
$\left|V_{CKM}^{corr}(i,j)\right|$ from
$\left|V_{CKM}^{tree}(i,j)\right|$ (for $i,j=1,2$) turns out to be
similar in size for CKM-ruled (see eq. \ref{minfv-CKM}) and
democratic (see eq. \ref{demfv-CKM}) textures. It is smallest for
the hierarchical texture (see eq. \ref{hierfv-CKM}). Therefore,
CKM-ruled texture in (\ref{minfv-yukawa}) and democratic one in
(\ref{demfv-yukawa}) exhibit a pronounced sensitivity to
supersymmetric threshold corrections in comparison to hierarchical
texture in (\ref{hierfv-yukawa}).

The physical quark fields, which arise after the unitary rotations
(\ref{diag-yukawa}), acquire the masses
\begin{eqnarray}
\overline{\bf M_{u}}(M_{weak})= \mbox{diag.}\left(0.055, 1.27, 144.78\right)\,,\; \overline{\bf
M_{d}}(M_{weak})= \mbox{diag.}\left(0.099, 0.27, 2.4\right)
\end{eqnarray}
all measured in ${\rm GeV}$. In this physical basis for quark
fields, $V_{CKM}^{corr}$ is responsible for charged current
interactions in the effective theory below $M_{weak}$. The morale
of the analysis above is that, the high-scale flavor structures
(\ref{demfv-yukawa}) are to be modified in such a way that
$V_{CKM}^{corr}$ agrees with $V_{CKM}^{exp}$ with sufficient
precision. Aftermath, the question is to predict quark masses
appropriately at $Q=M_{weak}$ so that, depending on the particle
spectrum of the effective theory beneath, existing experimental
values of quark masses at $Q=M_Z$ are reproduced correctly.

\subsection{Inclusion of flavor violation from squark soft masses}
In this section we extend GUT-scale flavor structures analyzed in
Sec. 3.1 by switching on flavor mixings in certain squark soft
mass-squareds. In other words, we maintain Yukawa textures to be
one of (\ref{minfv-yukawa}), (\ref{hierfv-yukawa}) or
(\ref{demfv-yukawa}), and examine what happens to CKM prediction
if squared masses of squarks possess non-trivial flavor mixings at
the GUT scale.

The effective Yukawa couplings ${\bf Y_{u,d}}^{eff}$ beneath
$Q=M_{weak}$ receive contributions from all entries of ${\bf
m_{Q,U,D}^2}(M_{weak})$ via respective mass insertions
\cite{higgs}. Generically, larger the mass insertions larger the
flavor violation potential of ${\bf Y_{u,d}}^{eff}$. Consequently,
main problem is to determine the relative strengths of on-diagonal
and off-diagonal entries of  ${\bf m_{Q,U,D}^2}(M_{weak})$  given
that they start with a certain pattern of flavor mixings. Take,
for instance, ${\bf m_{Q}^2}$ which evolves with energy scale via
(\ref{rg-mq}) at single loop level. Its analytic solution is
difficult, if not impossible, given coupled nature of RGEs and
further complications brought about by the presence of flavor
mixings. However, for an approximate yet instructive analysis, one
can consider solving (\ref{rg-mq}) for an infinitesimally small
scale change from $M_{GUT}$ down to $M_{GUT}-\Delta Q$:
\begin{eqnarray}
{\bf m_{Q}^2}(\Delta t)&=& {\bf m_{Q}^2}(0) - \frac{61}{99} \Delta
t\, m_{1/2}^2 {\bf 1}\nonumber\\ &+& \Delta t \left\{{\bf
m_{Q}^2}(0) + 2 m_0^2 {\bf 1}, {\bf Y_u}^{\dagger} {\bf Y_u} +
{\bf Y_d}^{\dagger} {\bf Y_d}\right\} + 2 \Delta t \left( {\bf
Y_u^A}^{\dagger} {\bf Y_u^A} + {\bf Y_d^A}^{\dagger} {\bf Y_d^A}
\right)
\end{eqnarray}
with the scale variable $\Delta t = (4 \pi)^{-2} \log (1- \Delta
Q/M_{GUT})$. Here right-handed squark mass-squareds are taken
strictly flavor-diagonal, for simplicity. This approximate
solution gives enough clue that diagonal entries of ${\bf
m_{Q}^2}$ tend to take hierarchically large values at the IR due
to the gluino mass contribution, mainly. However, its off-diagonal
entries do not have such an enhancement source:
\begin{eqnarray}
\label{off-diag} \left({\bf m_{Q}^2}\right)_{i j}(\Delta t)&=&
\left({\bf m_{Q}^2}\right)_{i j}(0) + \Delta t \left[ \left({\bf
m_{Q}^2}\right)_{i i}(0) + \left({\bf m_{Q}^2}\right)_{j j}(0) + 4
m_0^2 \right] \left({\bf Y_u}^{\dagger} {\bf Y_u} + {\bf
Y_d}^{\dagger}
{\bf Y_d}\right)_{i j}\nonumber\\
&+& \Delta t \left[\left({\bf Y_u}^{\dagger} {\bf Y_u} + {\bf
Y_d}^{\dagger} {\bf Y_d}\right)_{i i} + \left({\bf Y_u}^{\dagger}
{\bf Y_u} + {\bf Y_d}^{\dagger} {\bf Y_d}\right)_{j j} \right]
\left({\bf m_{Q}^2}\right)_{i j}(0)\nonumber\\
&+& \Delta t \left({\bf Y_u}^{\dagger} {\bf Y_u} + {\bf
Y_d}^{\dagger} {\bf Y_d}\right)_{6-(i+j)\, j} \left({\bf
m_{Q}^2}\right)_{i\, 6-(i+j)}(0)\nonumber\\ &+& \Delta t
\left({\bf Y_u}^{\dagger} {\bf Y_u} + {\bf Y_d}^{\dagger} {\bf
Y_d}\right)_{i\, 6-(i+j) } \left({\bf m_{Q}^2}\right)_{
6-(i+j)\, j}(0) \nonumber\\
&+& 2 \Delta t \left( {\bf Y_u^A}^{\dagger} {\bf Y_u^A} + {\bf
Y_d^A}^{\dagger} {\bf Y_d^A} \right)_{i j}
\end{eqnarray}
unless Yukawas or trilinear couplings are given appropriate
boundary configurations at the GUT scale. That this is the case
can be seen explicitly by considering, for instance,  democratic
texture for Yukawas (\ref{demfv-yukawa}) together with
(\ref{YApropY}) and strict universality and flavor-diagonality of
the soft masses, except
\begin{eqnarray}
\label{dem-mq} {\bf m_{Q}^2}(0) = m_0^2 \left( \begin{array}{ccc}
  1 & 1 & 1 \\
  1 & 1 & 1 \\
  1 & 1 & 1
\end{array} \right)
\end{eqnarray}
which contributes maximally to each term of (\ref{off-diag}). Even
with such a democratic pattern for Yukawas, trilinear couplings
and ${\bf m_{Q}^2}(0)$, however, one obtains at $M_{weak}=1\ {\rm
TeV}$
\begin{eqnarray}
 {\bf m_{Q}^2}= (533.37\ {\rm GeV})^2 \left(
\begin{array}{ccc}
  1.0 & -0.0512 & -0.0510 \\
  -0.0512 & 1.0 & -0.0513 \\
  -0.0510 & -0.0513 & 1.0
\end{array} \right)
\end{eqnarray}
with similar structures for ${\bf m_{U}^2}$ and ${\bf m_{D}^2}$.
Alternatively, if one adopts (\ref{minfv-yukawa}) or
(\ref{hierfv-yukawa}) setups  the off-diagonal entries of squark
soft mass-squareds at $M_{weak}$ are found to remain around
$m_0^2$ which are much smaller than the on-diagonal ones.
Therefore, Yukawa textures (and hence those of the trilinear
couplings) studied in Sec. 3.1 lead one generically to hierarchic
textures for squark soft mass-squareds at $Q=M_{weak}$
irrespective of how large the flavor mixings in ${\bf
m_{Q,U,D}^2}(0)$ might be. In fact, predictions for CKM matrix
remain rather close to those in Sec. 3.1 above.  This is actually
clear from (\ref{off-diag}) where off-diagonal entries of ${\bf
m_{Q,U,D}^2}$ are seen to evolve into new mixing patterns via
themselves and those of Yukawas and trilinear couplings. In
conclusion, evolution of squark soft masses is fundamentally
Yukawa-ruled and when Yukawas at the GUT scale are taken to shoot
the measured value of CKM matrix, the mass insertions associated
with  ${\bf m_{Q,U,D}^2}(M_{weak})$ are too small to give any
significant contribution to ${\bf Y_{u,d}}^{eff}$.

As follows from (\ref{off-diag}), for generating sizeable
off-diagonal entries for ${\bf m_{Q,U,D}^2}(M_{weak})$ it is
necessary to abandon either Yukawa textures analyzed in Sec. 3.1.
or proportionality of trilinear couplings with Yukawas. Therefore,
we take Yukawa couplings at the GUT scale precisely as
(\ref{demfv-yukawa}), we maintain (\ref{YApropY}) for both ${\bf
Y_d^{A}}$ and ${\bf Y_e^A}$, and we take ${\bf m_{U}^2}(0)$ and
${\bf m_{D}^2}(0)$ strictly flavor-diagonal as in all three case
studies carried out in Sec. 3.1. However, we take ${\bf
m_{Q}^2}(0)$ as in (\ref{dem-mq}) above, and ${\bf Y_u^A}$ as
\begin{eqnarray}
\label{dem-yua} {\bf Y_u^A}(0) = - 150\ {\rm GeV} \left(
\begin{array}{ccc}
  1 & 1 & 1 \\
  1 & 1 & 1 \\
  1 & 1 & 1
\end{array} \right)
\end{eqnarray}
which certainly violates (\ref{YApropY}) that enforces trilinears
to be proportional to the corresponding Yukawas. Then two-loop RG
running from $Q=M_{GUT}$ down to $Q=M_{weak}$ gives
\begin{eqnarray}
\label{xdemfv-YA} {\bf Y}_{u}^A&=& \left( \begin{array}{c c c}\matrix{ \ -262.087 & -259.342 & -260.709 \cr \
-259.474& -261.954 & -260.709\cr \ -260.688 & -260.674 & - 260.735 \cr } \
\end{array}\right)\nonumber\\
{\bf Y}_{d}^A&=&\left( \begin{array}{c c c}\matrix{\ -41.435 & 37.091\ e^{-0.0127 i} & - 40.408\ e^{0.0124 i}
\cr -  36.171\ e^{0.0102 i} & 52.230\ e^{-0.0019 i} & - 41.558\ e^{0.0228 i} \cr 39.983\ e^{-0.0300 i} &
42.075\ e^{-0.0224 i} & - 41.683\ e^{0.0025 i} \cr }\
\end{array}\right)~\end{eqnarray}
both measured in ${\rm GeV}$ at $M_{weak}=1\ {\rm TeV}$. Though
not shown explicitly, each entry of ${\bf Y_u^A}$ is complex with
a phase around $10^{-7}$ -- $10^{-6}$ in size. On the other hand,
squark soft mass-squared at $Q=M_{weak}$ are given by
\begin{eqnarray}
\label{xdemfv-msq} {\bf m_Q^2}&=& \left(516.58\ {\rm
GeV}\right)^{2} \left(\begin{array}{ccc}
  1.0 & - 0.13 & - 0.13 \\
  -0.13 & 1.0 &  -0.13\\
 -0.13 & -0.13 &1.0
\end{array}\right)
\nonumber\\
{\bf m_U^2}&=& \left(455.49\ {\rm GeV}\right)^{2}
\left(\begin{array}{ccc}
  1.0 & -0.3852 & -0.3853 \\
  -0.3852& 1.0 &  -0.3853\\
  -0.3853 & -0.3853  &1.0
\end{array}\right)
\nonumber\\
{\bf m_D^2}&=& \left(532.91\ {\rm GeV}\right)^{2}
\left(\begin{array}{ccc}
  1.0 & 4.34\ 10^{-3}\ e^{-0.01 i}& - 4.15\ 10^{-3}\ e^{0.01i}\\
   4.34\ 10^{-3}\ e^{-0.01 i} & 1.0 & - 4.55\ 10^{-3}\ e^{0.02 i}\\
   - 4.15\ 10^{-3}\ e^{0.01i} & - 4.55\ 10^{-3}\ e^{0.02i} & 1.0
\end{array}\right)
 \end{eqnarray}
where small phases in off-diagonal entries of ${\bf m_Q^2}$ and
${\bf m_U^2}$ are neglected. A comparison with (\ref{demfv-msq})
reveals spectacular enhancements in mass insertions pertaining
${\bf m_Q^2}$ and ${\bf m_U^2}$.

The trilinear couplings (\ref{xdemfv-YA}) and squark mass-squareds
(\ref{xdemfv-msq}) give rise to non-trivial changes in flavor
structures of ${\bf Y_{u,d}}(M_{weak})$ by generating effective
Yukawas ${\bf Y_{u,d}}^{eff}$ beneath $Q=M_{weak}$. Then the CKM
matrix $V_{CKM}^{tree}$ obtained from ${\bf Y_{u,d}}(M_{weak})$
and $V_{CKM}^{corr}$ obtained from ${\bf Y_{u,d}}^{eff}$ compare
as:
\begin{eqnarray}
\label{xdemfv-CKM}  \begin{tabular}{|c|c|}
  \hline $\left|V_{CKM}^{tree}\right|$ & $\left|V_{CKM}^{corr}\right|$ \\ \hline
\end{tabular} =
\left(
\begin{array}{ccc}
  \begin{tabular}{|c|c|}
   \hline $0.9748$ & $0.9637$ \\ \hline
  \end{tabular} & \begin{tabular}{|c|c|}
    \hline 0.2229 & 0.2668 \\ \hline
  \end{tabular} & \begin{tabular}{|c|c|}
    \hline 0.0083 & 0.0080 \\ \hline
  \end{tabular} \\ \\
  \begin{tabular}{|c|c|}
    \hline 0.2229 & 0.2666 \\ \hline
  \end{tabular} & \begin{tabular}{|c|c|}
    \hline $0.9739$ & $0.9626$ \\ \hline
  \end{tabular} & \begin{tabular}{|c|c|}
   \hline 0.0421 & 0.0480 \\ \hline
  \end{tabular} \\ \\
  \begin{tabular}{|c|c|}
   \hline $0.0092$ & $0.0132$ \\ \hline
  \end{tabular} & \begin{tabular}{|c|c|}
    \hline $0.0419$ &  $0.0468$\\ \hline
  \end{tabular} & \begin{tabular}{|c|c|}
    \hline $0.99908$ & $0.99888$ \\ \hline
  \end{tabular}
\end{array}
\right)
\end{eqnarray}
where left (right) window of $\begin{tabular}{|c|c|}
 \hline  { }& { }\\ \hline
\end{tabular}$
in $(i,j)$-th entry refers to $\left|V_{CKM}^{tree}(i,j)\right|$ (
$\left|V_{CKM}^{corr}(i,j)\right|$). Obviously, $|V_{CKM}^{tree}|$
agrees very well with $\left|V_{CKM}^{exp}\right|$ as was the case
in (\ref{demfv-CKM}). This qualifies (\ref{demfv-yukawa}) to be
the correct high-scale texture given present experimental
determination of $V_{CKM}$ at $Q=M_{Z}$. However, implementation
of supersymmetric threshold corrections is seen to leave a big
impact on certain entries of the physical CKM matrix. Indeed,
$\left|V_{CKM}^{corr}(1,1)\right|$, $
\left|V_{CKM}^{corr}(1,2)\right|$, $
\left|V_{CKM}^{corr}(2,1)\right|$, $
\left|V_{CKM}^{corr}(2,2)\right|$, $
\left|V_{CKM}^{corr}(3,3)\right|$ turn out to have $6.06 \sigma$,
$23.99 \sigma$, $23.89 \sigma$, $ 26.52 \sigma$, $4.35  \sigma$
significance levels, respectively. These are to be contrasted with
standard deviations computed for (\ref{demfv-CKM}) in Sec. 3.1.3
above. Needless to say, these deviations are far beyond the
experimental sensitivities and thus supersymmetric threshold
corrections completely disqualify the flavor textures
(\ref{demfv-yukawa}) in a way different than (\ref{demfv-CKM}) due
to new structures (\ref{dem-mq}) and (\ref{dem-yua}).

Finally, physical quark fields, which arise after the unitary
rotations (\ref{diag-yukawa}), acquire the masses
\begin{eqnarray}
\overline{\bf M_{u}}(M_{weak})= \mbox{diag.}\left(0.138, 1.26,
143.3\right)\,,\; \overline{\bf M_{d}}(M_{weak})=
\mbox{diag.}\left(0.140, 0.304, 2.42\right)
\end{eqnarray}
all measured in ${\rm GeV}$. These mass predictions are close to
those obtained within democratic texture. As in all cases
discussed in Sec. 3.1. especially light quark masses fall outside
the existing experimental bounds, and choice of the correct
high-scale texture must reproduce both $V_{CKM}^{corr}$ and quark
masses in sufficient agreement with experiment.

\subsection{A purely soft CKM ?}
In Sec. 3.1 and 3.2 we have discussed how prediction for the CKM
matrix depends crucially on the inclusion of the supersymmetric
threshold corrections. This we did by negation $i.e.$ we have
taken certain Yukawa textures which are known to generate CKM
matrix correctly at tree level, and then included threshold
corrections to demonstrate how those the would-be viable flavor
structures get disqualified.

In this section we will do the opposite $i.e.$ we will take a
Yukawa texture which is known not to work at all, and incorporate
supersymmetric threshold corrections to show how it can become a
viable one, at least approximately. For sure, a highly interesting
limit would be to start with exactly diagonal Yukawas at the GUT
scale and generate CKM matrix beneath $M_{weak}$ via purely soft
flavor violation $i.e.$ flavor violation from sfermion soft
mass-squareds and trilinear couplings, alone. However, this limit
seems difficult to realize, at least for SPS1a$^\prime$ parameter
values, since it may require tuning of various parameters, in
particular, soft mass-squareds of Higgs and quark sectors
\cite{higgs}. Even if this is done by a fine-grained scan of the
parameter space, it will possibly cost a great deal of
fine-tuning. Indeed, threshold corrections depend on ratios of the
soft masses \cite{higgs}, and generating a specific entry of the
CKM matrix can require a judiciously arranged hierarchy among
various soft mass parameters -- a parameter region certainly away
from the SPS1a$^\prime$ point.

Therefore, we relax the constraint of strict diagonality and
consider instead GUT-scale Yukawa matrices with five texture
zeroes which are known to be completely unphysical as they cannot
induce the CKM matrix \cite{frits}. In fact, this kind of textures
has recently been found to arise from heterotic string
\cite{hetero} when the low-energy theory is constrained to be
minimal supersymmetric model \cite{stringy}. Consequently, we take
Yukawas at $Q=M_{GUT}$ to be
\begin{eqnarray}
\label{stfv-yukawa} {\bf Y_{u}}&=& \left(\begin{array}{ccc}
  0 & 9.249\ 10^{-5} & 1.428\ 10^{-3} \\
   1.307\ 10^{-3}& 0 & 0 \\
   0.4675 & 0 & 0
\end{array} \right)\nonumber\\
{\bf Y_{d}}&=& \left(
\begin{array}{ccc}
0 & 9.0\ 10^{-5} & 1.3\ 10^{-3} \\
   1.42\ 10^{-3}& 0 & 0 \\
   0.047 & 0 & 0
\end{array} \right)
\end{eqnarray}
with no flavor violation in the lepton sector: ${\bf Y_e}=
\mbox{diag.}\left(1.9\ 10^{-5}, 0.004, 0.071\right)$. Both ${\bf
Y_u}$ and ${\bf Y_d}$ are endowed with five texture zeroes, and
they precisely conform to the structures found in effective
theories coming from the heterotic string \cite{hetero}.

Besides, though left unspecified in \cite{hetero}, we take
sfermion mass-squareds strictly flavor-diagonal as in Sec. 3.1,
and let ${\bf Y_e^A}$ obey (\ref{YApropY}). For trilinear
couplings pertaining to squark sector we take
\begin{eqnarray}
\label{0stfv-YA} {\bf Y}_{u}^A(0)&=&\left(
\begin{array}{c c c}\matrix{ \ 0
& 0 & 0 \cr 0 & -30.469 & -74.029\cr \ 0 & -74.029 & -97.406 \cr }
\
\end{array}\right)\nonumber\\
{\bf Y}_{d}^A(0)&=&\left(
\begin{array}{c c c}\matrix{ \ 0
& 0 & 0 \cr 0 & -25.241 & -68.185\cr \ 0 & -67.545 & -63.990 \cr }
\
\end{array}\right)
\end{eqnarray}
both measured in ${\rm GeV}$. These trilinear couplings do not
obey (\ref{YApropY}); they are given completely independent flavor
structures, in particular, they exhibit ${\cal{O}}(1)$ mixing
between second and third generations. The first generation of
squarks is decoupled from the rest completely.

Two-loop RG running down to $Q=M_{weak}$ modifies GUT-scale
textures (\ref{0stfv-YA}) to give
\begin{eqnarray}
\label{stfv-YA} {\bf Y}_{u}^A&=&\left(
\begin{array}{c c c}\matrix{ \ 0
& -0.157 & -2.426 \cr -1.326 & -75.382 & -183.335\cr \ -474.410 & -126.247 & -167.265 \cr } \
\end{array}\right) \nonumber\\
{\bf Y}_{d}^A&=&\left(
\begin{array}{c c c}\matrix{ \ 0
& -0.231 & -3.341 \cr -3.114 & -78.521 & -212.328\cr \ -103.062 &
-205.742 & -193.530 \cr } \
\end{array}\right)
\end{eqnarray}
both measured in ${\rm GeV}$. The texture zeroes in
(\ref{0stfv-YA}) are seen to elevated to small yet nonzero values
via RG running. The squark soft mass-squareds, on the other hand,
exhibit the following flavor structures at $M_{weak}=1\ {\rm
TeV}$:
\begin{eqnarray}
\label{stfv-msq} {\bf m_Q^2}&=& \left(560.63\ {\rm GeV}\right)^{2} \left(\begin{array}{ccc}
  0.936 & - 0.029 & - 0.036 \\
  -0.029 & 1.051 &  -0.049\\
 -0.036 & -0.049 & 1.012
\end{array}\right)
\nonumber\\
{\bf m_U^2}&=& \left(523.88\ {\rm GeV}\right)^{2} \left(\begin{array}{ccc}
  1.155 & -3.1\ 10^{-4} & -2.9\ 10^{-4} \\
  -3.1\ 10^{-4}& 1.107 &  -5.5\ 10^{-2}\\
  -2.9\ 10^{-4} & -5.5\ 10^{-2} &0.738
\end{array}\right)
\nonumber\\
{\bf m_D^2}&=& \left(548.52\ {\rm GeV}\right)^{2} \left(\begin{array}{ccc}
  1.043 & -3.72\ 10^{-4}& - 3.54\ 10^{-4}\\
   -3.72\ 10^{-4} & 0.997 & - 5.322 \ 10^{-2} \\
    - 3.54\ 10^{-4} &  - 5.322 \ 10^{-2} & 0.960
\end{array}\right)
 \end{eqnarray}
where off-diagonal entries are seen to be hierarchically small so
that contributions to ${\bf Y_{u,d}}^{eff}$ from squark soft
mass-squareds are expected to be rather small.

The use of Yukawas, trilinear couplings and squark mass-squareds,
all rescaled to $M_{weak}=1\ {\rm TeV}$ via RG running, give rise
to modifications in Yukawa couplings after squarks being
integrated out. In fact, the CKM matrix $V_{CKM}^{tree}$ obtained
from ${\bf Y_{u,d}}(M_{weak})$ and $V_{CKM}^{corr}$ obtained from
${\bf Y_{u,d}}^{eff}$ compare as:
\begin{eqnarray}
\label{stfv-CKM}  \begin{tabular}{|c|c|}
  \hline $\left|V_{CKM}^{tree}\right|$ & $\left|V_{CKM}^{corr}\right|$ \\ \hline
\end{tabular} =
\left(
\begin{array}{ccc}
  \begin{tabular}{|c|c|}
   \hline $0.9999$ & $0.9751$ \\ \hline
  \end{tabular} & \begin{tabular}{|c|c|}
    \hline 0.0044 & 0.2216 \\ \hline
  \end{tabular} & \begin{tabular}{|c|c|}
    \hline 0.0 & 0.0079\\ \hline
  \end{tabular} \\ \\
  \begin{tabular}{|c|c|}
    \hline 0.0044 & 0.2218 \\ \hline
  \end{tabular} & \begin{tabular}{|c|c|}
    \hline $0.9999$ & $0.9742$ \\ \hline
  \end{tabular} & \begin{tabular}{|c|c|}
   \hline 0.0 & 0.0412 \\ \hline
  \end{tabular} \\ \\
  \begin{tabular}{|c|c|}
   \hline $0.0$ & $0.0014$ \\ \hline
  \end{tabular} & \begin{tabular}{|c|c|}
    \hline $0.0$ &  $0.0419$\\ \hline
  \end{tabular} & \begin{tabular}{|c|c|}
    \hline $1.0$ & $0.99912$ \\ \hline
  \end{tabular}
\end{array}
\right)
\end{eqnarray}
where left (right) window of $\begin{tabular}{|c|c|}
 \hline  { }& { }\\ \hline
\end{tabular}$
in $(i,j)$-th entry refers to $\left|V_{CKM}^{tree}(i,j)\right|$ (
$\left|V_{CKM}^{corr}(i,j)\right|$).

It is clear that $V_{CKM}^{tree}$ by no means qualifies to be a
realistic CKM matrix: $\left|V_{CKM}^{tree}(i,j)\right| = 0$ for
$(i,j)=(1,3), (3,1), (2,3), (3,2)$; moreover, Cabibbo angle is
predicted to be one order of magnitude smaller. In addition, its
diagonal elements turn out to be well outside the experimental
limits. However, once supersymmetric threshold corrections are
included certain entries are found to attain their experimentally
preferred ranges. Indeed, $\left|V_{CKM}^{tree}(1,1)\right|$ and
$\left|V_{CKM}^{tree}(3,1)\right|$ fall right at their upper
bounds, and $\left|V_{CKM}^{tree}(1,3)\right|$ far exceeds the
experimental bound. The predictions for these entries are not good
enough; they need to be correctly predicted by further
arrangements of the GUT-scale textures. Nevertheless, for the main
purpose of illustrating how threshold corrections influence flavor
structures at the IR end,  the results above are good enough for
what has to be shown since all other entries turn out to be in
rather good agreement with experimental bounds. The case study
illustrated here shows that, even unphysical Yukawa textures with
five texture zeroes, can lead to acceptable CKM matrix predictions
once supersymmetric threshold corrections are incorporated into
Yukawa couplings.

The corrected Yukawa couplings lead to the following quark mass
spectrum:
\begin{eqnarray}
\overline{\bf M_{u}}(M_{weak})= \mbox{diag.}\left(0.168, 0.93,
151.6\right)\,,\; \overline{\bf M_{d}}(M_{weak})=
\mbox{diag.}\left(0.0325, 0.0711, 2.31\right)
\end{eqnarray}
all measured in ${\rm GeV}$. These predictions are not violatively
outside the experimental limits, except for the up quark mass. A
rehabilitated choice for the GUT-scale textures
(\ref{stfv-yukawa}) should lead to a fully consistent prediction
for CKM matrix (with much  better precision than in, especially
the  $(1,3)$, $(3,1)$ entries of (\ref{stfv-CKM}) above) together
with precise predictions for quark masses (modulo sizeable QCD
corrections while running from $Q=M_{weak}$ down to hadronic
scale).

\section{Conclusion}
In this work we have studied impact of integrating  superpartners
out of the spectrum on Yukawa couplings beneath the supersymmetry
breaking scale. In Sec. 2 we have outlined the formalism. In Sec.
3.1 we have illustrated effects of threshold corrections with
respect to certain high-scale Yukawa textures which are known, at
tree level, to lead to experimentally acceptable CKM predictions
at ${\rm TeV}$ scale. In Sec. 3.2 we have switched on flavor
violation in squark soft mass-squareds to determine their effects
on CKM prediction, and pointed out how important the flavor
structure of the trilinear couplings at the GUT scale for squark
soft masses to develop sizeable off-diagonal entries. In Sec. 3.3
above our approach was opposite to those in Sec. 3.1 and 3.2 in
that we have investigated how threshold corrections can
rehabilitate an unacceptable GUT-scale texture by using Yukawa
couplings with five texture zeroes. Our main conclusion is that
supersymmetric threshold corrections leave observable impact on
Yukawa couplings of quarks; confrontation of high-scale textures
with experiments at $Q=M_{Z}$ should take into account such
corrections.

We have focussed mainly on SPS1a$^{\prime}$ point so as to
standardize our predictions for various low-scale couplings and
masses. Though not shown explicitly, predictions for Higgs boson
masses as well as various other sparticle masses and couplings
turn out to be in good agreement with experimental bounds (as a
characteristic of SPS1a$^{\prime}$ point they should agree with
laboratory and astrophysical bounds modulo small variations in
certain parameters stemming from presence of the flavor violation
sources). For supersymmetric parameter regions outside
SPS1a$^{\prime}$, certain parameters are expected to vary
significantly, especially at large $\tan\beta$. However, scanning
of such parameter regions will result mainly in strengthening of
the statements arrived at rather conservative SPS1a$^{\prime}$
point in that supersymmetric threshold corrections should be taken
into account in both bottom-up and top-down approaches to
supersymmetric flavor problem.

\section{Acknowledgements}
We are grateful to D. A. Demir for invaluable discussions. The work of L. S. was partially supported by
post-doctoral fellowship of the Scientific and Technical Research Council of Turkey.

\end{document}